# Novel 2D Silica Monolayers with Tetrahedral and Octahedral Configurations

Gaoxue Wang[1], G. C. Loh[1,2], Ravindra Pandey[1*], and Shashi P. Karna[3]

[1]Department of Physics, Michigan Technological University, Houghton, Michigan 49931, USA
[2]Institute of High Performance Computing, 1 Fusionopolis Way, #16-16 Connexis, Singapore 138632
[3]US Army Research Laboratory, Weapons and Materials Research Directorate, ATTN: RDRL-WM, Aberdeen Proving Ground, MD 21005-5069, U.S.A.

(January 7, 2015)

*Email: pandey@mtu.edu
    shashi.p.karna.civ@mail.mil

**Abstract**

Free-standing and well-ordered two-dimensional (2D) silica monolayers with tetrahedral (*T*-silica) and octahedral (*O*-silica) building blocks are found to be stable by first principles calculations; *T*-silica is formed by corner-sharing $SiO_4$ tetrahedrons in a rectangular network and *O*-silica consists of edge-sharing $SiO_6$ octahedrons. Moreover, the insulating *O*-silica is the strongest silica monolayer, and can therefore act as a supporting substrate for nanostructures in sensing and catalytic applications. Nanoribbons of *T*-silica are metallic while those of *O*-silica have band gaps regardless of the chirality. We find the interaction of *O*-silica with graphene to be weak suggesting the possibility of its use as a monolayer dielectric material for graphene-based devices. Considering that the six-fold coordinated silica exists at high pressure in the bulk phase, the prediction of a small energy difference of *O*-silica with the synthesized silica bilayer together with the thermal stability at 1000 K suggest that synthesis of *O*-silica can be achieved in experiments.

Silica ($SiO_2$) is one of the most abundant materials in Earth's crust. Bulk silica has a rich phase diagram which includes *α*- and *β*-quartz, *α*- and *β*-cristobalite, tridymite, and stishovite [1]. In general, silica prefers $SiO_4$ tetrahedral unit at ambient conditions consisting of fourfold-coordinated Si at the center and twofold-coordinated O at the corners [1,2]. At high pressure, the phases formed by $SiO_6$ octahedral unit with sixfold-coordinated Si at the center and threefold-coordinated O at the corners are observed [3-6]. Stishovite silica with the octahedral unit is thermodynamically stable above 7 GPa [4,5,7], and is the densest among the allotropes of silica due to its compact structure.

Silica films spontaneously form on a clean silicon surface when the surface is exposed to air. Such two-dimensional (2D) silica films have found their importance in several applications, such as dielectric layer in integrated circuits and supporting substrate for catalysis [8]. However, their amorphous nature limits the dielectric properties [9], and blocks the fabrication of continuous interfaces formed by the dielectric layer with other materials [10,11]. From an application's perspective, well-ordered and freestanding 2D silica will have great advantage over amorphous films. In particular, the monolayer crystalline silica is a much sought-after 2D material as an ideal dielectric for the van der Waals (vdW) heterostructures [10,12].

Following the stability criterion of the bulk form, 2D silica consisting of $SiO_4$ tetrahedrons has recently been investigated. Silicatene with the $SiO_{2.5}$ stoichiometry is realized on the Mo (112) surface [8,13]. It is formed by corner-sharing $SiO_4$ tetrahedrons in a hexagonal pattern. One of the oxygen atoms in each tetrahedron is bonded strongly to the Mo surface through the Si-O-Mo link, thus limiting the transferability of silicatene. Also, a bilayer silica has been successfully deposited on Ru(0001) surface with the corner-sharing tetrahedrons [14]. It has the $SiO_2$ stoichiometry, and is weakly anchored on the metal surface due to vdW interactions [9,14,15]. Such ultrathin bilayers, which could possibly be peeled off from the substrate is now being explored by experiments [14,16-18]. On the theoretical front, structural and electronic properties of *α*-silica with alternating $sp^3$ and $sp^2$ bonds has been reported recently [19]. In spite of the extensive experimental efforts, freestanding and transferable monolayer silica has not yet been realized and the monolayer configuration beyond the tetrahedral building blocks has seldom been reported [19-21]. There also appears to be a lack of fundamental understanding about the stability, electronic structure and suitability of silica monolayers as a non-interacting dielectric in van der Waals electronics.

Considering the importance of free-standing 2D oxides as potential dielectric layers in nanoelectronic devices [10,22] and a wide range of other applications, such as photonics and optoelectronics, we have investigated the possible free-standing silica monolayers together with their stability, electronic band structure, and mechanical properties with the use of PSO and DFT. We have also investigated the electronic band structure of silica "nanoribbons" with armchair and zig-zag edges. For the 2D silica, we have calculated and reported the scanning-tunneling microscope (STM) images. Our calculations show that the 2D silica monolayer with the $SiO_4$ tetrahedral structure in a rectangular network (*T*-silica) is stable. More importantly, we find that a 2D silica with the octahedral configuration (*O*-silica) is also stable, and is more compact and stronger than the previously reported monolayers including *α*-silica and silicatene [13,19]. Finally, in order to assess the suitability of the 2D silica in nanoelctronics, we have calculated the equilibrium configuration, stability and the interaction between two *O*-Silica monolayers and a graphene monolayer and an *O*-silica monolayer. Our calculations show a rather weak (van der Waals type) interaction between graphene and *O*-silica monolayers, suggesting the applicability of the latter as a dielectric in graphene-based nanoelectronics.

The stable configurations of the 2D silica structures were obtained using the PSO methodology as implemented in the CALYPSO program package [23-26]. Constraining the stoichiometry to be $SiO_2$, the given configuration is optimized which, in the next cycle, (i.e. generation) surmounts an energy barrier to arrive at a lower-energy configuration following the PSO method. In our calculations, the number of configurations (e.g., population) that is produced at each step was set to 24, and the number of CALYPSO steps (e.g., generation) was fixed to 30. The history of structural search in terms of energy of each structure in each generation finds 2D configurations with rectangular network (i.e. *T*-silica) and octahedral building block (i.e. *O*-silica) as the energetically preferred monolayer configurations besides other configurations with complex silica network (See supplementary material, Figures S1, S2 and S3).

The energetically preferred configurations were then fully relaxed to determine their equilibrium structural and electronic properties. The norm-conserving Troullier-Martins pseudopotential [27] and the Perdew-Burke-Ernzerhof (PBE) [28] exchange correlation functional form of the DFT together with a double-zeta plus polarization function basis sets were used. The energy convergence was set to $10^{-5}$ eV. The mesh cutoff energy was chosen to be 500 Ry. The geometry optimization was considered to be converged when the residual force on each atom was smaller than 0.01 eV/Å. The reciprocal space was sampled by a grid of (11×11×1) *k* points in the Brillouin zone. In the periodic supercell approach, the vacuum distance normal to the plane was larger than 20 Å to eliminate the interaction between the replicas. The phonon dispersion calculation was based on Vibra of SIESTA utility [29].

To benchmark our modeling elements, we first considered the cases of *α*-silica ($Si_2O_3$) monolayer and silicatene ($Si_2O_5$) for calculations. Our calculated bond lengths of 1.83 Å (for $sp^3$ bond) and 1.62 Å (for $sp^2$ bond) for *α*-silica are in reasonably good agreement with the corresponding values of 1.76 Å and 1.58 Å obtained at previously reported using PBE-DFT level of theory [19] (See supplementary material, Figure S4). Our calculated short and long $R_{Si-O}$ bonds for silicatene are calculated to be 1.65 Å and 1.72 Å, respectively. The calculated lattice constant of 5.23 Å for silicatene is in excellent agreement with the corresponding experimental value of 5.2-5.4 Å [13]. Note that $R_{Si-O}$ in the bulk *α*-quartz is reported to be 1.61 Å [30].

Next, monolayers of *T*- and *O*-silica were considered to determine their energetic stabilities relative to the silica bilayer, which has been synthesized recently [14]. Figure 1 shows the variation of the formation energy with area per Si atom for silica bilayer, *O*-silica and *T*-silica. The formation energy of the 2D silica configuration with respect to its constituent atoms can is calculated as

$$E_{form} = (E_{total} - N_{Si}E_{Si} - N_O E_O)/(N_{Si} + N_O) \qquad (1),$$

where, $E_{total}$ is the total energy of the system. $E_{Si}$, and $E_O$ are the atomic energies of Si and O which are calculated to be -123.3 and -440.2 eV, respectively. $N_{Si}$ and $N_O$ are the number of Si and O atoms in the unit cell. The values of ($N_{Si}$, $N_O$) are (1, 2), (1, 2), and (4, 8) for *T*-, *O*-, and silica bilayer, respectively. A negative formation energy suggests the configuration to be stable or metastable configuration [31]. Note that the energy minimum in each curve of Figure 1 represents the equilibrium configuration of the given structure.

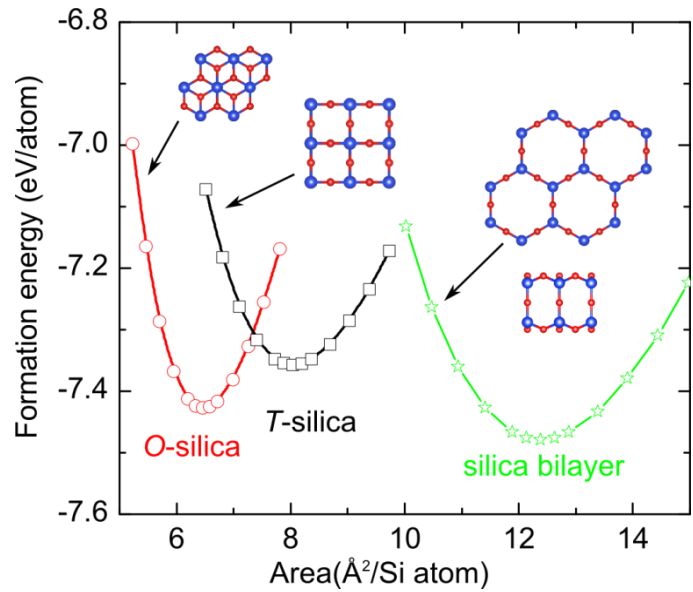

*Figure 1. The calculated formation energy vs. area/Si atom for the 2D silica configurations.*

The calculated formation energies of *O*- and *T*-silica are -7.43, and -7.36 eV/atom, respectively (Table 1). Interestingly, these values are very close to the value of the formation energy of -7.48 eV/atom for the silica bilayer which has recently been synthesized [14]. Our results therefore suggest that these novel silica monolayers can be synthesized. In comparison, the calculated formation energies of *α*-silica and silicatene are -6.66 and -6.85 eV/atom, respectively (Figure S5).

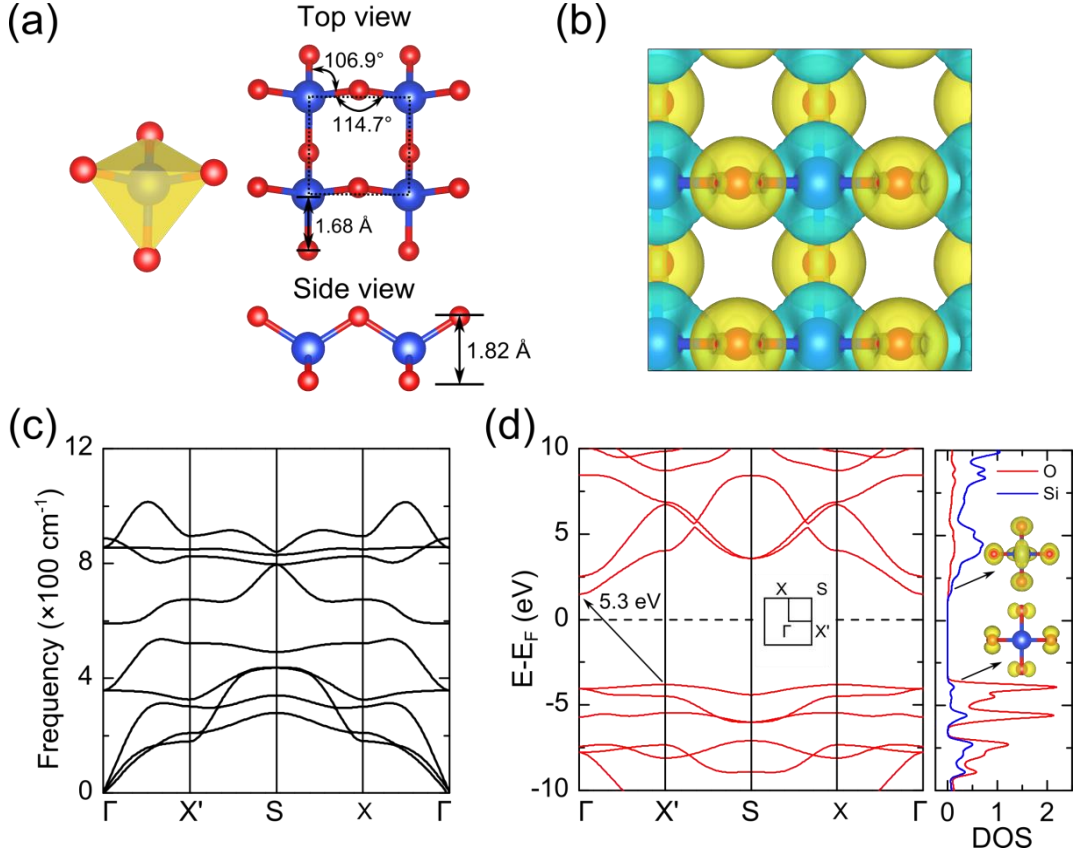

*Figure 2. T-silica. (a) atomic structure, (b) deformation electron density (with isosurface of 0.01 e/Å³). Blue represents depletion of electrons, and yellow represents accumulation of electrons, (c) phonon dispersion curves, and (d) electronic band structure and DOS. Inset show the charge density at VBM and CBM. The high symmetry points are Γ(0,0,0), X' (1/2,0,0), S(1/2,1/2,0) and X (0,1/2,0).*

*T*-silica is constituted by $SiO_4$ tetrahedrons in a rectangular network (Figure 2). The corner O atom is shared by the neighboring tetrahedrons. $R_{Si-O}$ is 1.68 Å and ∠O-Si-O is 106.9° in the equilibrium configuration of *T*-silica. Note that ∠O-Si-O for a perfect tetrahedron is 109.5°. Absence of imaginary vibrational frequency in the calculated phonon dispersion curves (Figure 2(c)) implies that *T*-silica is dynamically stable. The speed of sound is ~1.9 km/s obtained by the slope of the longitudinal acoustic (LA) branch near Γ. The maximum vibration frequency of the optical branch is 1015 cm$^{-1}$. The deformation charge density plot shown in Figure 2 (b) suggests the charge transfer from Si to O yielding to the charge of Si to be 3.52e as per Mulliken charge analysis. Therefore, the Si-O bond retains partial ionic character in *T*-silica. The valence band maximum (VBM) at X' is dominated by O-*p* orbitals and the conduction band minimum (CBM) at Γ is composed of Si-*s* and O-*p* orbitals as shown by the orbital-projected DOS in Figure 2(d). *T*-silica has an indirect (X'→ Γ) band gap of 5.3 eV whereas the direct gap at Γ is 5.6 eV.

Unlike silica bilayer and *T*-silica, the basic building blocks of *O*-silica are the SiO$_6$ octahedron as shown in Figure 3. Each Si is surrounded by six O atoms forming an octahedron. *O*-silica has the structure similar to that of the metallic 1T phase of MoS$_2$ [32]; Si layer is sandwiched between the top and bottom O layers. It also has the appearance of a centered honeycomb (top view of Figure 3(a)). Our calculated $R_{Si-O}$ is 1.84 Å, which is slightly larger than the measured value of axial Si-O bond length of 1.81 Å in stishovite [33]. The ∠O-Si-O bond angles are calculated to be 83.9° and 96.1°, which are distorted from those (≈90°) in a perfect octahedron. Each O is threefold-coordinated, with three Si-O bonds not lying in the same plane, and ∠Si-O-Si is 96.1°.

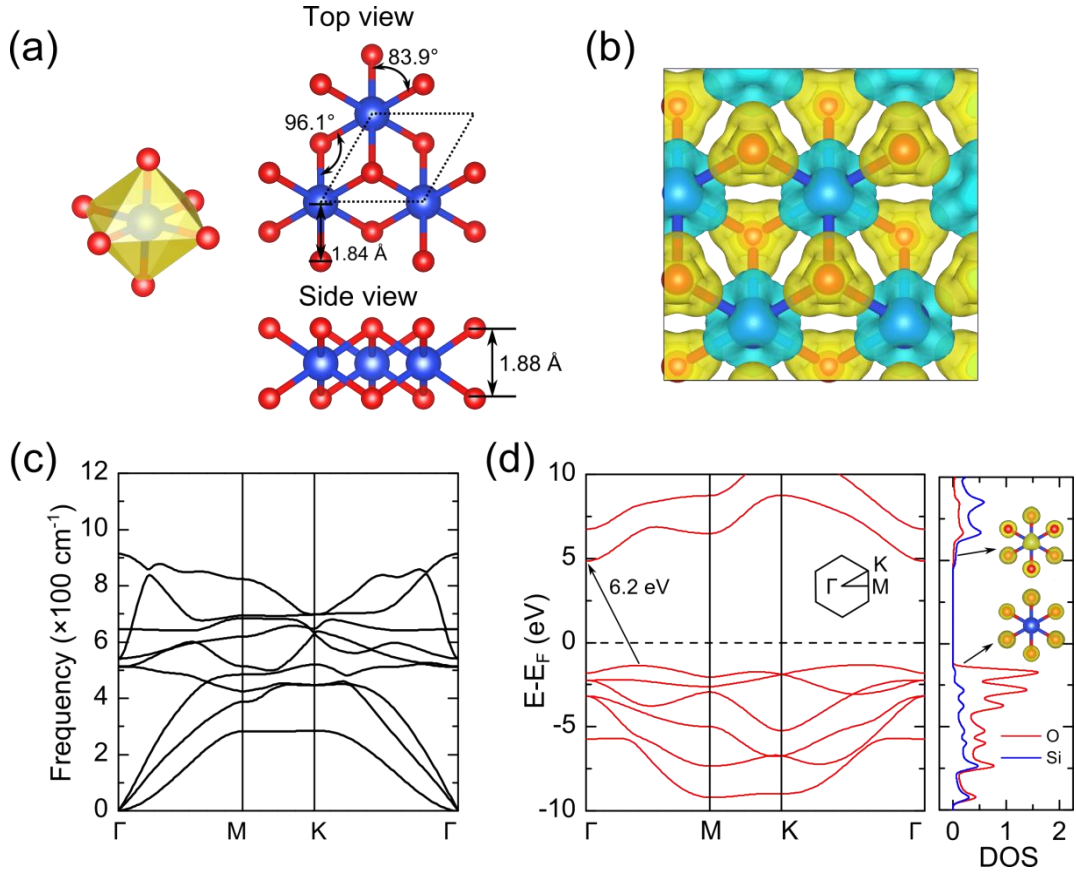

*Figure 3. O-silica. (a) atomic structure, (b) deformation electron density (with isosurface of 0.01 e/Å³). Blue represents depletion of electrons, and yellow represents accumulation of electrons, (c) phonon dispersion curves, and (d) electronic band structure and DOS. Inset shows the charge density at VBM and CBM. The high symmetry points are Γ(0,0,0), K(2/3,1/3,0), and M (1/2,0,0).*

The phonon dispersion curves show the dynamical stability of *O*-silica (Figure 3(c)). The speed of sound is 2.3 km/s obtained from the slope of the LA braches near Γ. The maximum vibrational frequency of the optical branch is 912 cm$^{-1}$, which is smaller than that in *T*-silica, suggesting that the Si-O bond is slightly weaker in *O*-silica. This is consistent with the order of $R_{Si-O}$ in going from *T*-silica (1.68 Å) to *O*-silica (1.84 Å). For the bulk phase, $R_{Si-O}$ also increases in going from the tetrahedral phase to the high-pressure octahedral phase [2].

The deformation charge density distribution shows the sixfold-coordinated character of Si (Figure 3(b)). Mulliken charge analysis shows the Si charge to be 3.49e which is fractionally larger than that of fourfold-coordinated Si in *T*-silica. The VBM is dominated by O-*p* orbitals, and CBM is composed of both Si-*s* and O-*p* orbitals (Figure 3(d)). Note that the location of VBM is away from Γ. The calculated band gap is indirect (Γ-M → Γ) with a value of ~6.2 eV, and the direct band gap at Γ is 6.7 eV.

The thermal stability of *O*- and *T*-silica is further examined by *ab initio* molecular dynamics (MD) simulations. The MD simulations were performed at 1000 K with a time step of 1 *fs* using Nośe heat bath scheme [34]. The averages of the total energies for the structures remain nearly constant during the simulation, and the structures remain unchanged after 5 *ps* (See supplementary materials, Figure S8), which suggests *O*- and *T*-silica are thermally stable.

Table 1. Structural properties of 2D silica structures. $a_0$ is the lattice constant, $R_{Si-O}$ is the near-neighbor distance, $R_{Si-Si}$ is the distance between Si atoms, $E_{form}$ is the formation energy defined by Eq. 1, and C is the in-plane stiffness. For silica bilayer, the experimental values are given in parenthesis.

| This work | $a_0$ (Å) | $R_{Si-O}$ (Å) | $R_{Si-Si}$ (Å) | $E_{form}$ (eV/atom) | C (N/m) |
|---|---|---|---|---|---|
| *T*-silica (SiO$_2$) | 2.84 | 1.68 | 2.84 | -7.36 | 159 |
| *O*-silica (SiO$_2$) | 2.73 | 1.84 | 2.73 | -7.43 | 215 |
| Silica bilayer (Si$_4$O$_8$) | 5.35 (5.40)[16] | 1.65 (1.67)[18] | 3.09 (3.12)[18] | -7.48 | 154 |

In order to understand and establish the mechanical properties of 2D silica structures, we calculated the in-plane stiffness [35-37], $C = \frac{1}{S_0}(\frac{\partial^2 E_S}{\partial \varepsilon^2})$. Here, $S_0$ is the equilibrium area of the structure, $E_S$ is the strain energy, defined as the energy difference between the strained and the relaxed structure, and $\varepsilon$ is the uniaxial strain applied to the structure. In order to establish the reliability of our approach, we first calculated the in-plane stiffness for graphene. The calculated value *C* for graphene is 342 N/m, which is in excellent agreement with the corresponding experimental value of 340 ± 40 N/m [38], giving confidence in our calculated values of *C* for 2D silica.

Our calculations show that *O*-silica has the largest in-plane stiffness of 215 N/m among the 2D silica structures (Table 1). This is in line with *O*-silica being more compact than other 2D silica structures (Figure 1). These results, therefore, suggest that *O*-silica could be freestanding and a good support for applications in catalysis and electronic devices [39]. In contrast, the stiffness constant for *α*-silica and silicatene is calculated to be 17 and 45 N/m, respectively, which are much smaller than that for the *O*-silica, suggesting the former two to be relatively soft compared to the latter.

To get further insight into the stability of six-fold coordinated silica monolayer, the terms contributing to the total energy are compared in Table S1 of the supplementary material. A smaller $R_{Si-Si}$ leads to the highest ion-ion Coulomb interaction in *O*-silica. However, the compact electronic arrangement of Si and O atoms yields the ion-electron Coulomb energy to be the lowest in *O*-silica. This interplay of bonding and geometrical arrangements facilitates the overall comparable total energy for *T*-, *O*-silica and silica bilayer. Analysis of the electron density with the quantum theory of "atoms in molecule" via the electron localization function (ELF), charge density and Laplacian plots shows that the Si–O bond is slightly ionic in nature in octahedral coordinated silica compared to that in tetrahedral coordinated silica layers (See supplementary materials Figure S6 and S7).

Next, we performed simulation of the scanning tunneling microscopy (STM) images based on Tersoff and Hamann approximation [40] at constant current mode to mimic the experimental setup [41-44]. A gold cluster, $Au_{13}$ was used to represent the cap configuration of the STM tip. The calculated STM images are shown in Figure 4. The simulated STM image of the bilayer silica (Figure 4(d)) is in excellent agreement with the corresponding measured STM image [18], showing surface with a hexagonal pattern. The simulated STM image of the *T*-silica (Figure 4(a)), on the other hand, shows a stripe-like surface feature. *O*-silica has triangle-like surface character (Figure 4(b)) contributed by the top O atoms, while *α*-silica shows flat dot-like surface (Figure 4(c)) due to the $sp^2$ bonding features. We expect that the distinct characteristics of 2D silica structures will be useful in identifying and distinguishing future experimental STM images of these nano-scale materials.

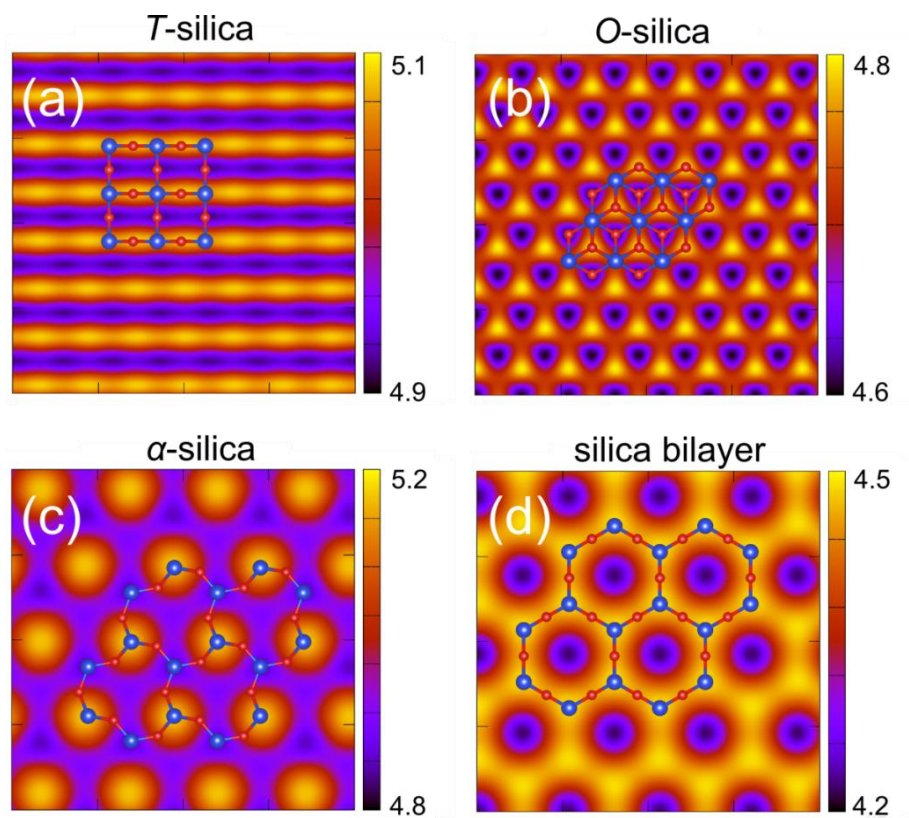

*Figure 4. Simulated STM images for 2D silica using constant current mode. The size of the STM images are 20 Å×20 Å, the scale bar are in the units of angstrom.*

Integration of the 2D materials into electronic devices can generally be performed by cutting the 2D sheet into finite components, such as nanoribbons. For *T*-silica, cutting can be done along the rectangular lattice direction (See supplementary material, Figure S9(a)). In the equilibrium configuration, we find that atoms near the edges of nanoribbon are largely reconstructed, and the ribbon is curved laterally with ∠O-Si-O to 109.2° which is very close to that of a perfect octahedron. *T*-silica nanoribbon is metallic and the energy band crossing the Fermi level is derived from dangling bonds of the edge Si atoms. For *O*-silica, cutting along the armchair direction leads to only one kind of edge as shown in the inset of Figure S9(b). If we cut along the zigzag direction, two kinds of edges terminated by O atoms or Si atoms appeared for the zigzag nanoribbons. The nanoribbons are stable with minor reconstruction near the edge in their equilibrium configurations. The states associated with the dangling bonds at the edges do not cross the Fermi level, and the band gap is independent of the edge configurations of the *O*-silica zigzag nanoribbons. (The structural details of *T*- and *O*-silica nanoribbons are provided in the supplementary materials, Figures S9, S10, and S11)

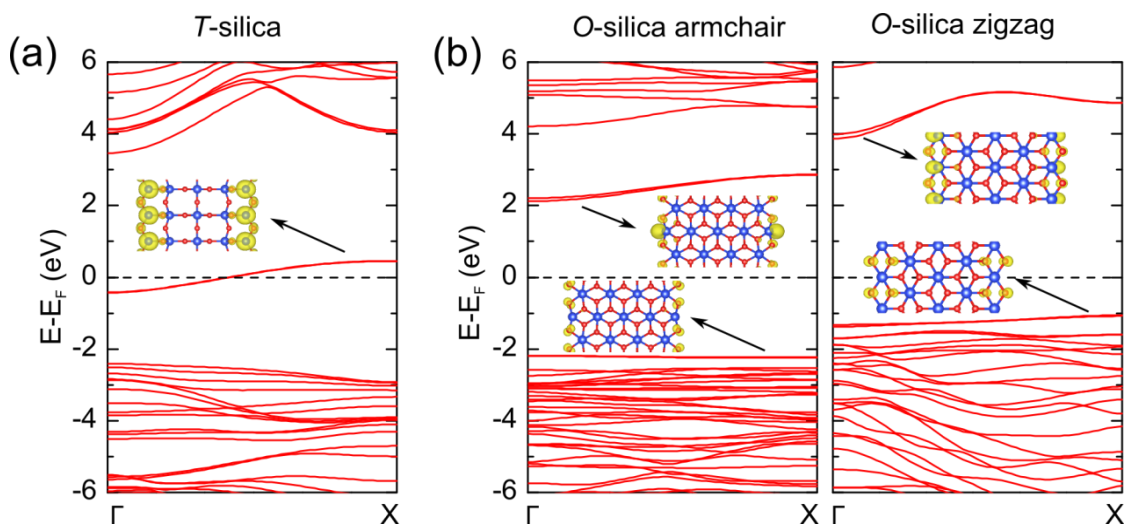

*Figure 5. Nanoribbons - Band structure of (a) T-silica nanoribbon and (b) O-silica nanoribbon in the armchair and ziagzag configurations. The insets show charge density associated with edge states.*

Finally, in order to address the feasibility of *O*-silica as a dielectric monolayer in the vdW heterostructures, e.g. graphene-based devices, we performed electronic structure calculations, on the graphene/*O*-silica and the *O*-silica/*O*-silica bilayers in the configuration shown in Figure 6. The graphene/*O*-silica heterostructure (Figure 6) was simulated by a supercell of $(2 \times 2)\boldsymbol{R}$ for graphene and $(\sqrt{3} \times \sqrt{3})\boldsymbol{R}$ for *O*-silica minimizing the lattice mismatch to be about ~3%.

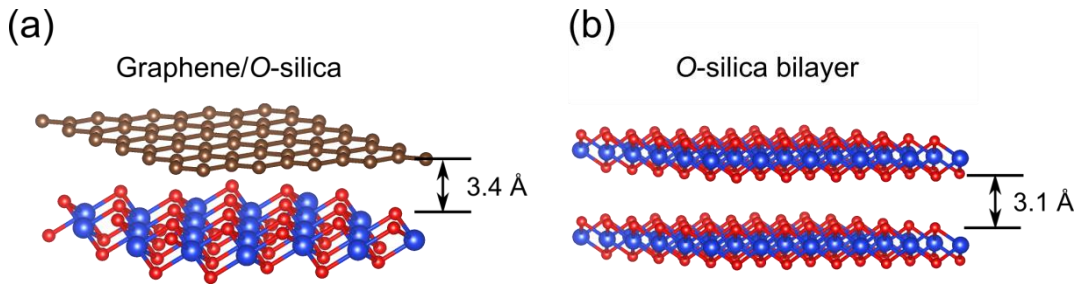

*Figure 6. The equilibrium configurations of (a) the graphene/O-silica heterostructure and (b) O-silica bilayer.*

The vdW correction to the correlation-exchange functional was considered for the heterostructure [45]. To benchmark the bilayer system, we first considered the bilayer *O*-silica for which the interlayer distance is calculated to be ~3.1 Å with the band gap of 5.6 eV. For graphene/*O*-silica bilayer, the calculated interlayer distance is ~3.4 Å and the binding energy with respect to its constituent monolayers is 24 meV/atom. The results therefore suggest the interaction between graphene and *O*-silica to be weak leading to retention of the electronic characteristics of graphene in the heterostructure (See supplementary materials, Figure S12). Interestingly, the band gap in the heterostructure does not open up when we reduce the interlayer distance to ~2.0 Å, thus demonstrating that *O*-silica could potentially act as an ideal monolayer insulator in the so-called vdW heterostructure devices.

In summary, from our first-principles DFT calculations, silica sheets constituted by either edge-sharing $SiO_6$ octahedrons (i.e. *O*-silica) or rectangular network of $SiO_4$ tetrahedrons (i.e. *T*-silica) are found to be stable. Both 2D silica structures have large in-plane stiffness (mechanical strength) and electrical insulating properties. Additionally, *O*-silica nanoribbons also have band gaps which are not dependent on their edge configurations, whereas *T*-silica nanoribbons are metallic. The structural simplicity and stiffness thus facilitate the use of freestanding *O*- and *T*-silica monolayers for versatile applications, e.g. as a single layer insulator in vdW heterostructures or as a supportive substrate for catalysis.

*O*-silica is the first proposed 2D silica with octahedral-coordinated building blocks, which is of fundamental importance considering that six-fold coordinated silica in the bulk phase exists at high pressure. It has been reported that the high-pressure phases of cubic $TiO_2$ and rocksalt GaN could be stabilized in thin films [46,47]. Also, 2D materials with similar structure, e.g., $MoS_2$ [32] has been successfully synthesized via molecular beam epitaxy (MBE) and chemical vapor deposition (CVD) methods. We expect that our results showing a small energy difference of *O*-silica with the synthesized silica bilayer would stimulate and help experimental research to realize *O*-silica.

**Acknowledgements**

Helpful discussions with S. Gowtham are acknowledged. RAMA and Superior, high performance computing clusters at Michigan Technological University, were used in obtaining results presented in this paper. The authors appreciate D. R. Banyai for providing the code for STM simulation. Financial support from ARL W911NF-14-2-0088 is acknowledged.

**Supplementary materials**

**Novel 2D Silica Monolayers with Tetrahedral and Octahedral Configurations**
Gaoxue Wang, G. C. Loh, Ravindra Pandey, and Shashi P. Karna

## *1. Structure search by the particle swarm optimization (PSO) method (CALYPSO program package)*

The structural search was performed by using the particle swarm optimization (PSO) method as implemented in CALYPSO code [1-4]. PSO is a method for multidimensional optimization, which is inspired by the social behavior of birds flocking [2]. It can predict the crystal structure with the known chemical composition at given external conditions, which has successfully utilized for the structure prediction of lithium-boron compounds at high pressure [5] and other 2D materials [6]. In our case, the structures with the $SiO_2$ stoichiometry are considered for calculations. The number of structures (e.g., population) that produced at each step is set to 24, and the number of CALYPSO steps (e.g., generation) is fixed to 30. The required structural relaxation by CALYPSO was performed using the Perdew-Burke-Ernzerhof (PBE) [7] exchange correlation functional to density functional theory (DFT) [3].

Figures S1, S2, and S3 show the history of structural search by CALYPSO. For the unit cell with 3 atoms, only *O*-, and *T*-silica are predicted as the energetically preferred monolayer as shown in Figure S1.

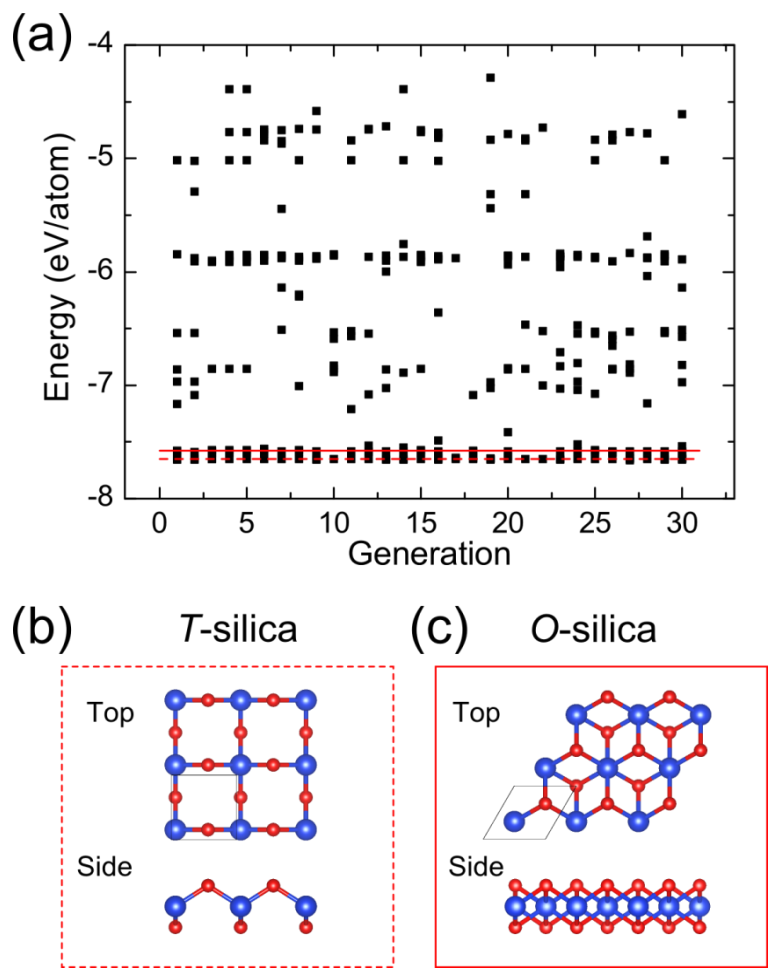

*Figure S1*. *(a) The history of structural search by CALYPSO for the unit cell with 3 atoms (Si:O=1:2). O-, and T-silica are predicted as the energetically preferable structures as illustrated by the red lines. (b) and (c) are the structural configurations of T- an O-silica, respectively.*

For the unit cell with 6 atoms, other stable structures are also obtained besides *T*- and *O*-silica. *T'*-silica can be obtained by flipping 1/2 of the SiO$_4$ tetrahedrons in *T*-silica, though its 'top' view is similar to that of T-silica. *T'*-silica has a distorted surface compared to *T*-silica, it will not therefore be further discussed in this paper. Other possible structure as shown in Figure S2(c) will not be considered for further calculations due to its structural complexity.

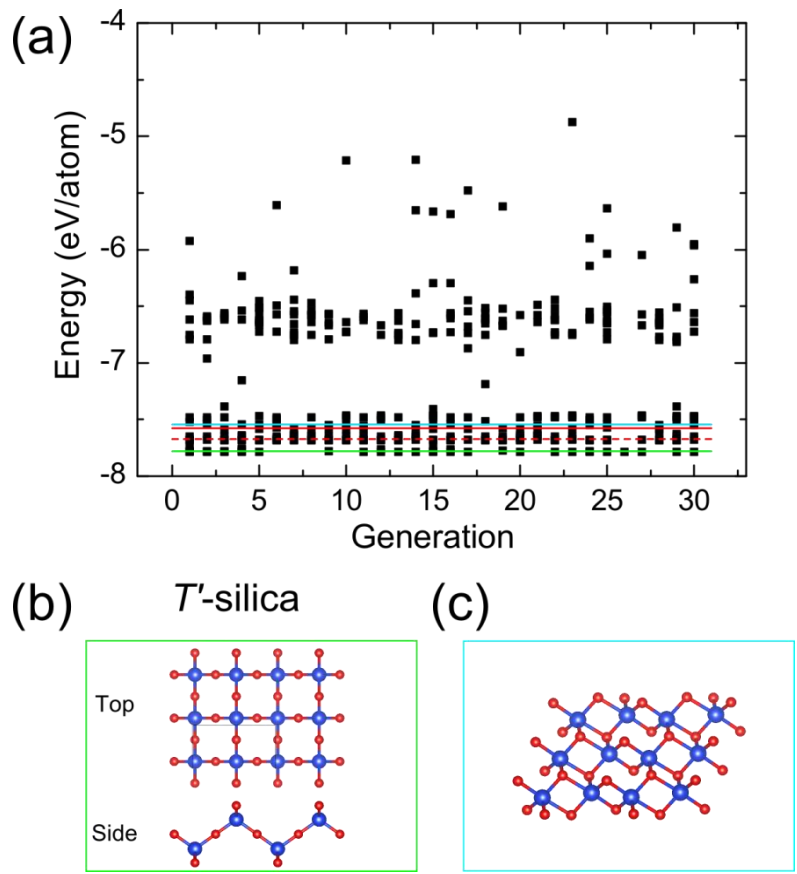

*Figure S2. (a) The history of structural search by CALYPSO for the unit cell with 6 atoms (Si:O=2:4). Red lines show T-, and O-silica, and other possible structuers are shown in (b) and (c).*

For the unit cell with 9 atoms, *O*-, *T*- and *T'*-silica are obtained together with the other structures with tetrahedral building blocks as shown in Figures S3(b) and S3(c).

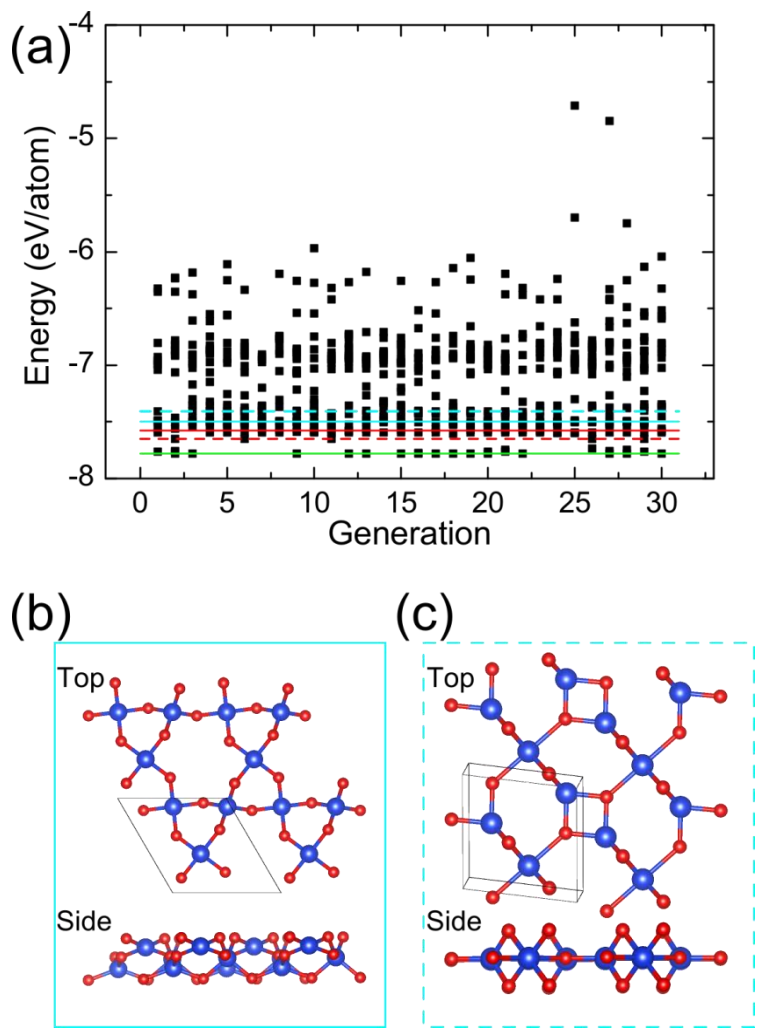

*Figure S3. (a) The history of structural search by CALYPSO for the unit cell with 9 atoms (Si:O=3:6). Red lines show T-, and O-silica, and green lines show T'-silica, other possible structuers are shown in (b) and (c).*

## 2. Structural and electronic properties of α-silica

In α-silica, each Si is threefold-coordinated either by coplanar O atoms (forming $sp^2$ bonds) or non-coplanar O atoms (forming $sp^3$ bonds). The $sp^3$ bonded Si atom occupies one corner of the tetrahedron, three O atoms occupy the other three corners, as seen from Figure S4(a). $R_{Si-O}$ of the $sp^3$ bonds is 1.83 Å, and $R_{Si-O}$ of $sp^2$ bonds is 1.62 Å. It is a semiconductor with the VBM contributed by $sp^3$ bonds, and CBM contributed by $sp^2$ bonds as seen in Figure 4(c).

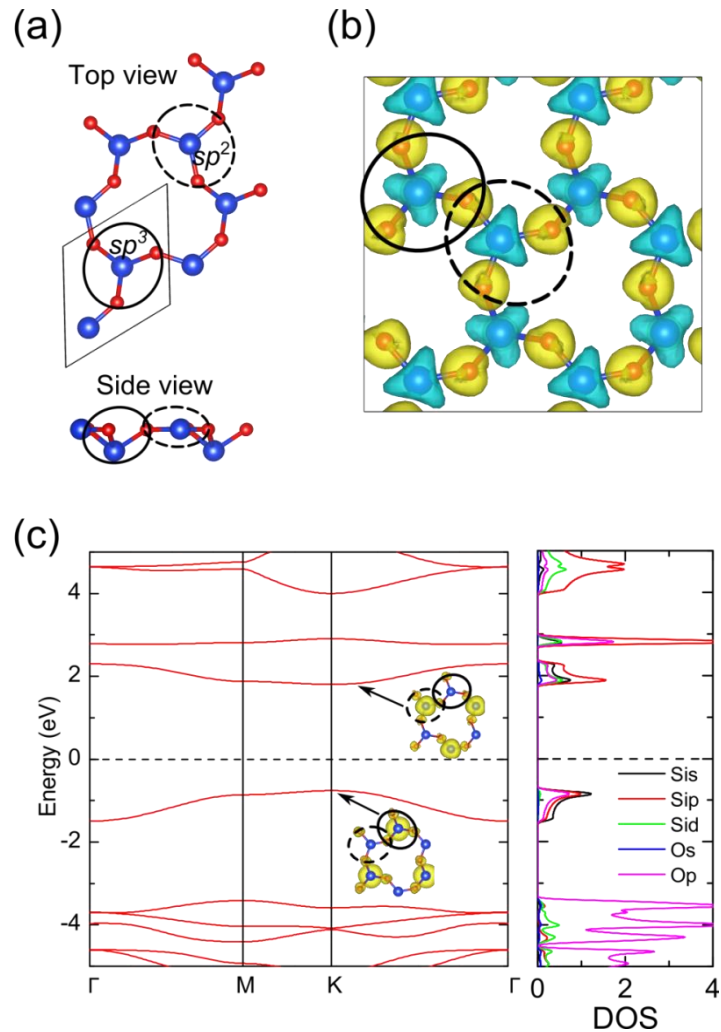

*Figure S4. α-silica. (a) Atomic structure where solid (dashed) circle illustrates the $sp^3$ ($sp^2$) bonds, (b) deformation electron density (with isosurface of 0.01 e/Å$^3$). Blue represents depletion of electrons, and yellow represents accumulation of electrons, (c) electronic band structure and DOS, inset shows the charge density at VBM and CBM.*

## 3. Formation energy of α-silica and silicatene

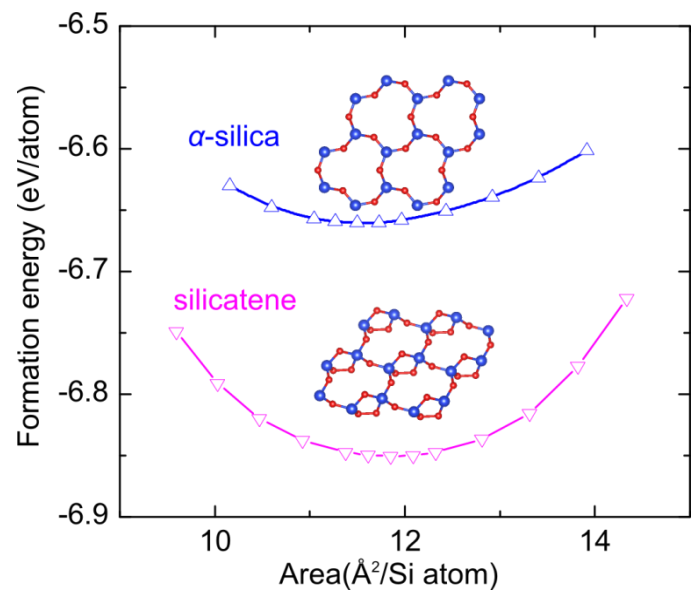

*Figure S5. Formation energy of α-silica and silicatene.*

## *4. QTAIM analysis*

The Si-O bond nature in the 2D silica allotropes is investigated by analyzing the topology of electron density with the quantum theory of atoms in molecule (QTAIM)[8,9]. The calculations were performed with the Perdew-Burke-Ernzerhof (PBE) functional form and the plane-wave basis sets as implemented in VASP.

Figure S6 shows the electron localization function (ELF) for 2D silica configurations. A large *ELF* around O atoms is mainly due to the difference of electronegativity between Si and O atoms. Tetrahedral configurations of silica bilayer and *T*-silica lead to similar ELF around the oxygen atom whereas the *ELF* of *O*-silica shows triangular shape reflecting the threefold-coordinated nature of O atoms.

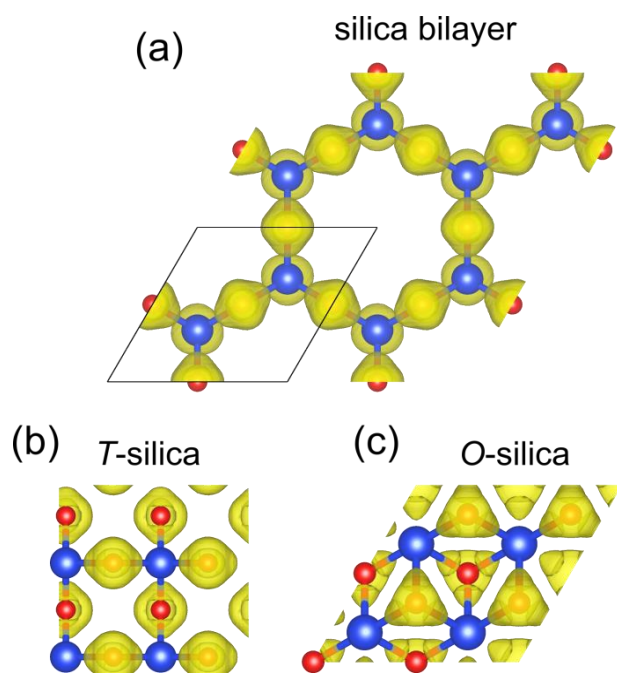

*Figure S6. Electron localization function (ELF) for bilayer silica, T- and O-silica with the isosurface at the isovalue of 0.8.*

The charge density and Laplacian plots (Figures S7(a) and S7(b)) also similarity in the nature of the Si-O bonds in silica bilayer and *T*-silica. In contrast, the Si-O bond in *O*-silica has more ionic characteristics as shown by the separated zero envelope of the 2D Laplacian (Figure S7(b)).

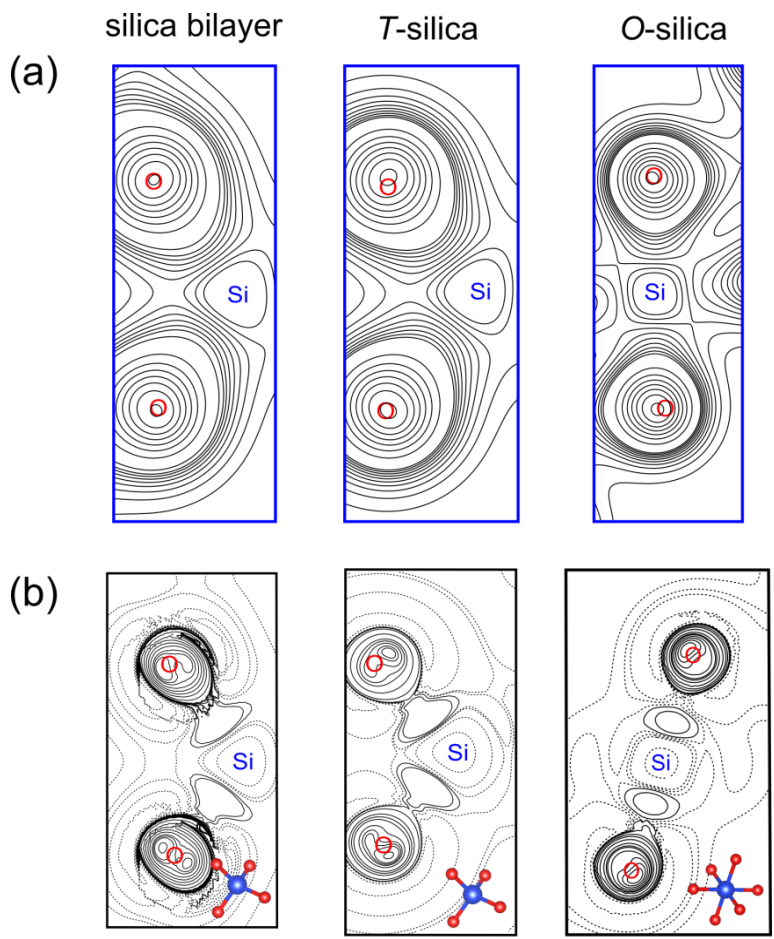

*Figure S7. 2D charge density contours (a) and Laplacian (b) for silica bilayer, T- and O-silica.*

## 5. Molecular dynamics (MD) simulation

The MD simulations are based on the Nośe thermostat. We used a large supercell of 5×5 to minimize the constraints induced by periodicity in the slab model. The time step is 1 *fs* up to 5 *ps*.

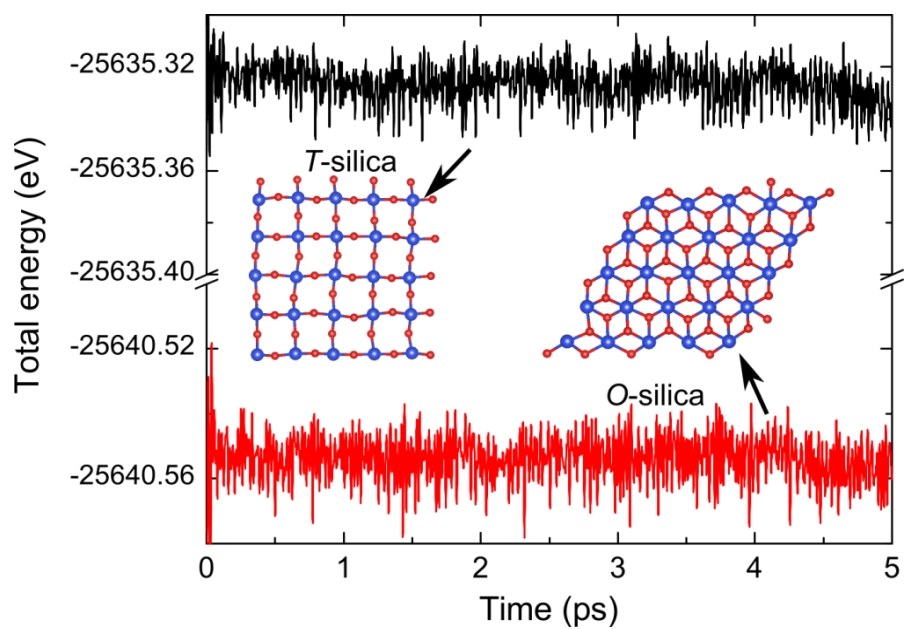

*Figure S8. Energy fluctuations of T- and O-silica as a function of simulation steps. The insets show the corresponding structueres with the simulation up to 5 ps.*

## *6. T- and O-silica nanoribbons*

For *T*-silica, we only considered the configuration by cutting along the line in which the edge is terminated by alternating Si and O atoms as illustrated by the shadowed region of Figure S9(a). For *O*-silica, cutting along the zigzag direction leads to edges terminated by Si (Si edge) or O atoms (O edge). Therefore, there exists three kinds of zigzag nanoribbons; (i) both edges terminated by O (O-O zigzag), (ii) both edges terminated by Si atoms (Si-Si zigzag), and (iii) one edge terminated by O and the other terminated by Si atoms (Si-O zigzag). On the other hand, cutting along the armchair direction yields the armchair nanoribbons with symmetric edges, or asymmetric edges (Figure S9(b)).

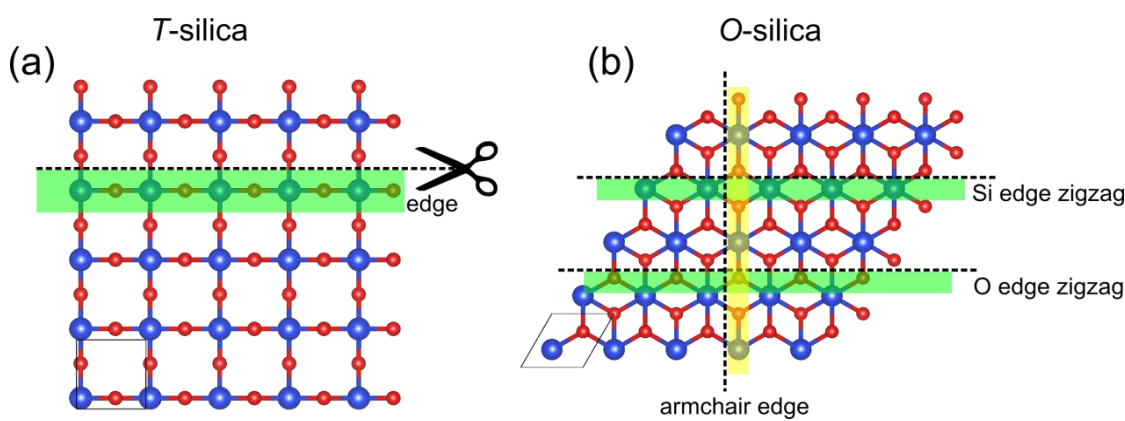

*Figure S9. T- and O-silica nanoribbons. The dashed line shows the cutting position, and the shadowed region shows the atoms at the edge of the nanoribbons.*

*T*-silica nanoribbons are predicted to be metallic as shown in Figure S10. The states crossing the Fermi level are derived from edge states, which are mainly contributed by the dangling bonds of Si atoms at the edges as seen from both the charge density distribution and DOS.

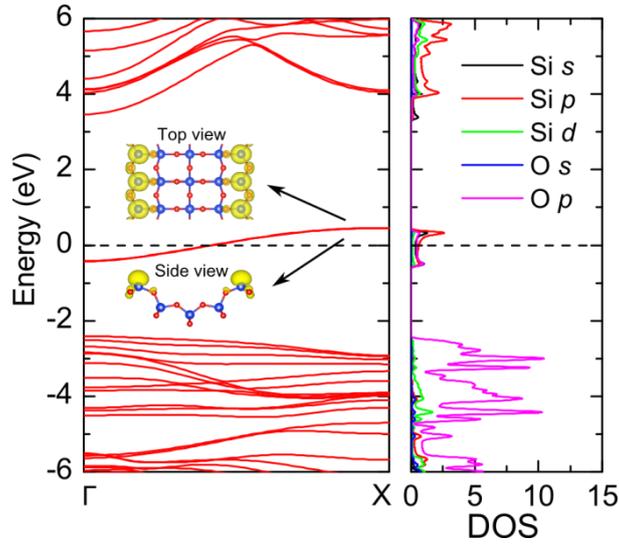

*Figure S10. Band structures and projected-DOS of T-silica nanoribbon.*

*O*-silica nanoribbons have the finite band gaps (Figure S11). The unsaturated edge Si atoms form VBM and CBM as seen from the charge density distribution and DOS.

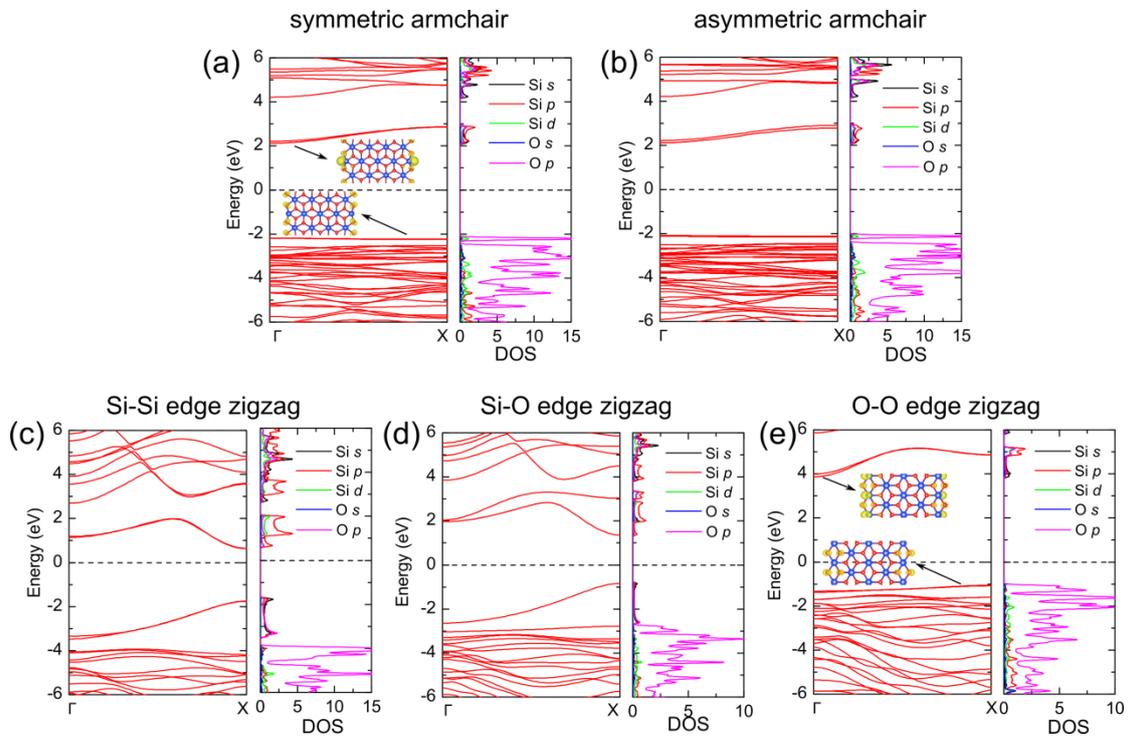

*Figure S11. Band structures and projected-DOS of O-silica nanoribbons with different edges.*

## 7. Band structure of Graphene/O-silica bilayer

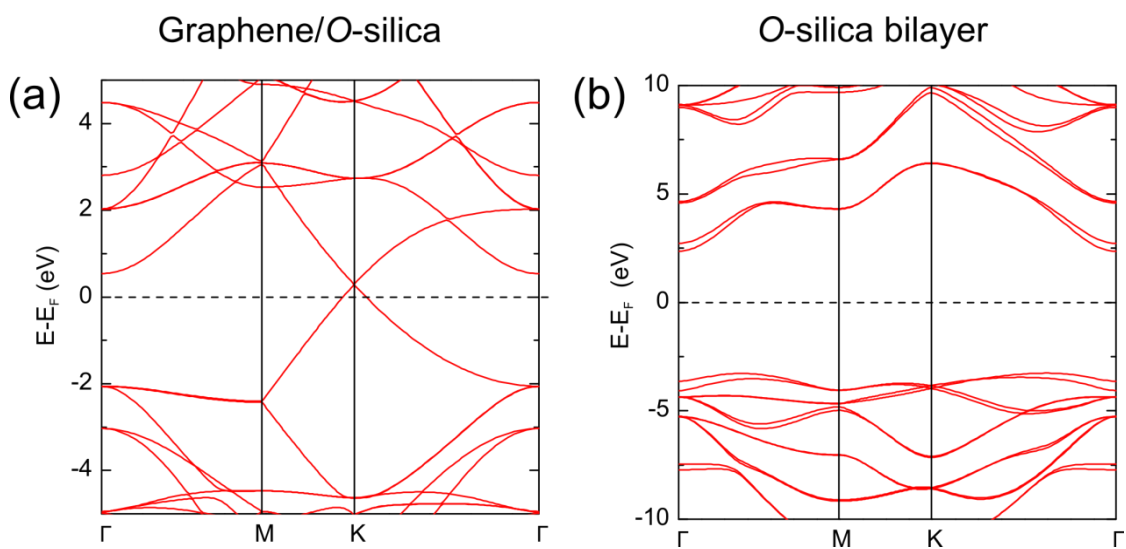

*Figure S12. Band structure of (a) $(2 \times 2)R$ Graphene/$(\sqrt{3} \times \sqrt{3})R$ O-silica heterostructure, and (b) O-silica bilayer. The Dirac cone of graphene is maintained in the graphene/O-silica bilayer.*

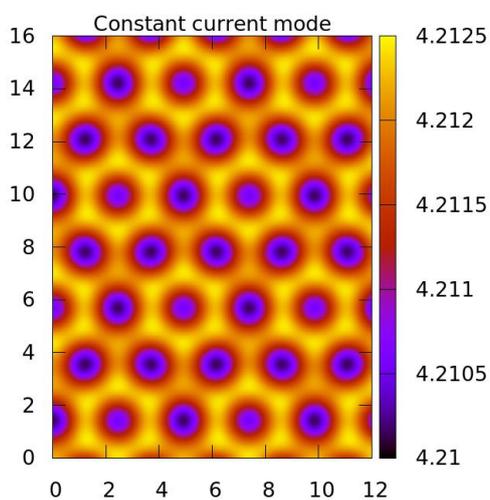

*Figure S13. Simulated STM image for Graphene/O-silica heterostructure.*

*Table S1*. *Decomposition of the total energy of T-, O-, and silica bilayer. The energy is in unit of eV/atom.*

|  | *T*-silica | *O*-silica | Silica bilayer |
|---|---|---|---|
| Kinetic | 217.49 | 218.01 | 219.40 |
| Hartree | 965.44 | 1094.34 | 376.41 |
| Exchange correlation | -93.27 | -93.72 | -93.42 |
| Ion-electron | -2129.99 | -2391.06 | -953.15 |
| Ion-ion | 698.40 | 830.44 | 108.71 |
| Total | -341.93 | -342.00 | -342.05 |